\documentclass[conference]{IEEEtran}
\IEEEoverridecommandlockouts
\usepackage[bottom=1.02in,top=0.71in,left=0.7in,right=0.7in]{geometry}
\usepackage{amsmath}
\usepackage{amsfonts}
\usepackage{amssymb}
\usepackage[ruled,vlined]{algorithm2e}
\usepackage{mathtools}  
\usepackage{amsthm}
\usepackage{booktabs}

\usepackage{tabulary}
\usepackage{multirow}
\usepackage{footnote}
\usepackage[export]{adjustbox}
\usepackage{booktabs}
\usepackage[justification=centering]{caption}
\usepackage{comment}

\usepackage{cases}
\usepackage{graphicx}
\usepackage{cite}
\usepackage{multicol}
\usepackage{xcolor}
\usepackage{graphicx}
\graphicspath{{./}{figures/}}
\newcommand\numeq[1]%
  {\stackrel{\scriptscriptstyle(\mkern-1.5mu#1\mkern-1.5mu)}{=}}
\usepackage{stfloats}% <-- added
\usepackage{cuted}
\usepackage{lipsum}  

\usepackage{dsfont}
\usepackage{amsmath,amssymb}

\usepackage{optidef}
\usepackage{amsmath}
\usepackage{caption}
\usepackage[font=footnotesize]{subfig}

\title{PHY-Layer Modeling and Throughput-Driven Adaptation for Batteryless V2X Networks}

\author{
    \IEEEauthorblockN{
        Zhaoyu Liu\IEEEauthorrefmark{1}, 
        Ruikang Li\IEEEauthorrefmark{1},
        Liu Cao\IEEEauthorrefmark{1},
        YuKun Pan\IEEEauthorrefmark{1},
        Xiangkai Wang\IEEEauthorrefmark{1},
        Lyutianyang Zhang\IEEEauthorrefmark{2}\\
    }
    \IEEEauthorblockA{
        \IEEEauthorrefmark{1}City University of Hong Kong (Dongguan), Dongguan, China\\
\IEEEauthorrefmark{2} The School of Microelectronics and Communication Engineering, Chongqing University, Chongqing, China\\
        Emails: \{72515198, 72515426, liu.cao, 72515531,72501108\}@cityu-dg.edu.cn, zhanglyutianyang@cqu.edu.cn
    }
    \vspace{-0.5cm}
\thanks{This work is supported by Guangdong and Hong Kong Universities ``1+1+1" Joint Research Collaboration Scheme - General Program.}
}

\begin{document}

\maketitle
\thispagestyle{empty}

\begin{abstract}
Passive overlay communication for batteryless devices is an important enabling capability for next-generation vehicle-to-everything (V2X) networks. However, enabling reliable passive payload delivery without occupying additional spectrum remains challenging, since overlay signaling must be embedded into short and time-varying vehicular packets while preserving the decodability of the legacy host transmission. This paper investigates a packetized batteryless V2X overlay architecture in which a dedicated short-range communications (DSRC)-based packet simultaneously carries conventional V2X data and a passive overlay payload. A compact PHY-layer model is developed to characterize the coupled effects of attenuation depth, embedded-bit rate, and legacy modulation and coding scheme (MCS) on host-link and passive-link reliability, as well as packet-level embedding feasibility. We then formulate a sum-throughput maximization problem that jointly accounts for the legacy packet error rate and passive decoding error rate. We further propose a multi-agent reinforcement learning (MARL)-based adaptive parameter-selection method. Simulation results show that the proposed MARL controller achieves stable convergence and improves the average throughput by 15\%, demonstrating the effectiveness of throughput-driven PHY adaptation for batteryless V2X overlay communications.
\end{abstract}


\section{Introduction}
With the rapid growth of vehicle-to-everything (V2X) networks, future intelligent transportation systems will require large numbers of ultra-low-cost passive endpoints, such as batteryless tags, roadside markers, and lightweight sensing nodes, to support infrastructure awareness and environmental monitoring \cite{clancy2024wireless_access_v2x,anwar2019phy_eval_v2x,abboud2016interworking_dsrc_cellular,moradi_pari2023dsrc_vs_lte_v2x,ansari2021joint_use_dsrc_cv2x}. Although recent backscatter and batteryless Internet of Things (IoT) techniques have significantly reduced the energy cost of uplink connectivity, reliable downlink delivery to such passive devices remains difficult because conventional receiver chains are still too power-hungry for batteryless operation \cite{jiang2023backscatter_survey_outlook,rezaei2023coding_backscatter_survey,zhang2026datadrivenbatterylesschannelsounding,cao2025lightweightcoordinateconditioneddiffusionapproach}.

A promising solution is to reuse already-occupied vehicular airtime instead of transmitting additional packets \cite{10757964}. In particular, Glaze showed that a preexisting wireless signal can be overlaid with an extra low-rate downlink through controlled amplitude attenuation while keeping the incumbent transmission decodable \cite{kapetanovic2019glaze,guo2022saiyan_lora_backscatter}. However, directly applying this idea to vehicular communications is difficult because vehicular packets are short, time-varying, and sensitive to PHY-layer parameter mismatch. In such a setting, the attenuation depth that improves passive detectability may simultaneously degrade legacy packet decoding, the overlay bit rate that increases passive throughput may reduce decoding robustness, and a higher host modulation and coding scheme (MCS) may improve nominal host throughput. Therefore, the design objective is not to maximize the passive rate alone, but to identify PHY operating points that balance host-link throughput and passive-link throughput.

Recent V2X studies mainly focus on link-reliability enhancement, congestion-aware transmission, and throughput optimization for the legacy vehicular link itself \cite{bianchi2000dcf,fu2024otcc_nr_v2x,11151706,kavas_torris2022cav_uav_v2x}. On the other hand, recent battery-free and backscatter studies investigate low-power architectures and coexistence mechanisms for passive communications \cite{jiang2023backscatter_survey_outlook,rezaei2023coding_backscatter_survey}. However, these two lines of work have not been tightly integrated for packetized vehicular overlay. In particular, the coupled effects of attenuation depth, overlay bit rate, and host MCS on legacy throughput, passive throughput, and packet-level embedding feasibility remain insufficiently understood in practical batteryless V2X systems.

Motivated by the aforementioned issues, we propose a packetized batteryless V2X overlay architecture at the PHY layer, where a legacy Wi-Fi/dedicated short-range communications (DSRC)-based packet serves as the host waveform and a passive payload is embedded through controlled amplitude attenuation. We formulate a constrained sum-throughput maximization problem under reliability and embedding-feasibility constraints, and employ multi-agent reinforcement learning (MARL) to adapt the attenuation depth, overlay bit rate, and host MCS based on historical look-up-table (LUT)-driven PHY evaluations. The main contributions of this paper are summarized as follows:
\begin{itemize}
\item We develop a PHY-layer model for packetized batteryless V2X overlay and analyze how the attenuation depth, embedded-bit rate, and legacy MCS jointly affect host-link decoding performance, passive-link decoding reliability, and embedding feasibility.
\item We formulate a sum-throughput maximization problem and propose an MARL-based parameter-adaptation method to jointly optimize the attenuation depth, embedded-bit rate, and legacy MCS, thereby improving the total delivered throughput under reliability constraints.
\end{itemize}

The rest of this paper is organized as follows. Section II presents the system architecture and system model. Section III formulates the optimization problem and introduces the proposed adaptation method. Section IV provides the simulation results and discussion. Section V concludes the paper.
\begin{figure*}[t]
    \centering
    \includegraphics[width=0.8\textwidth]{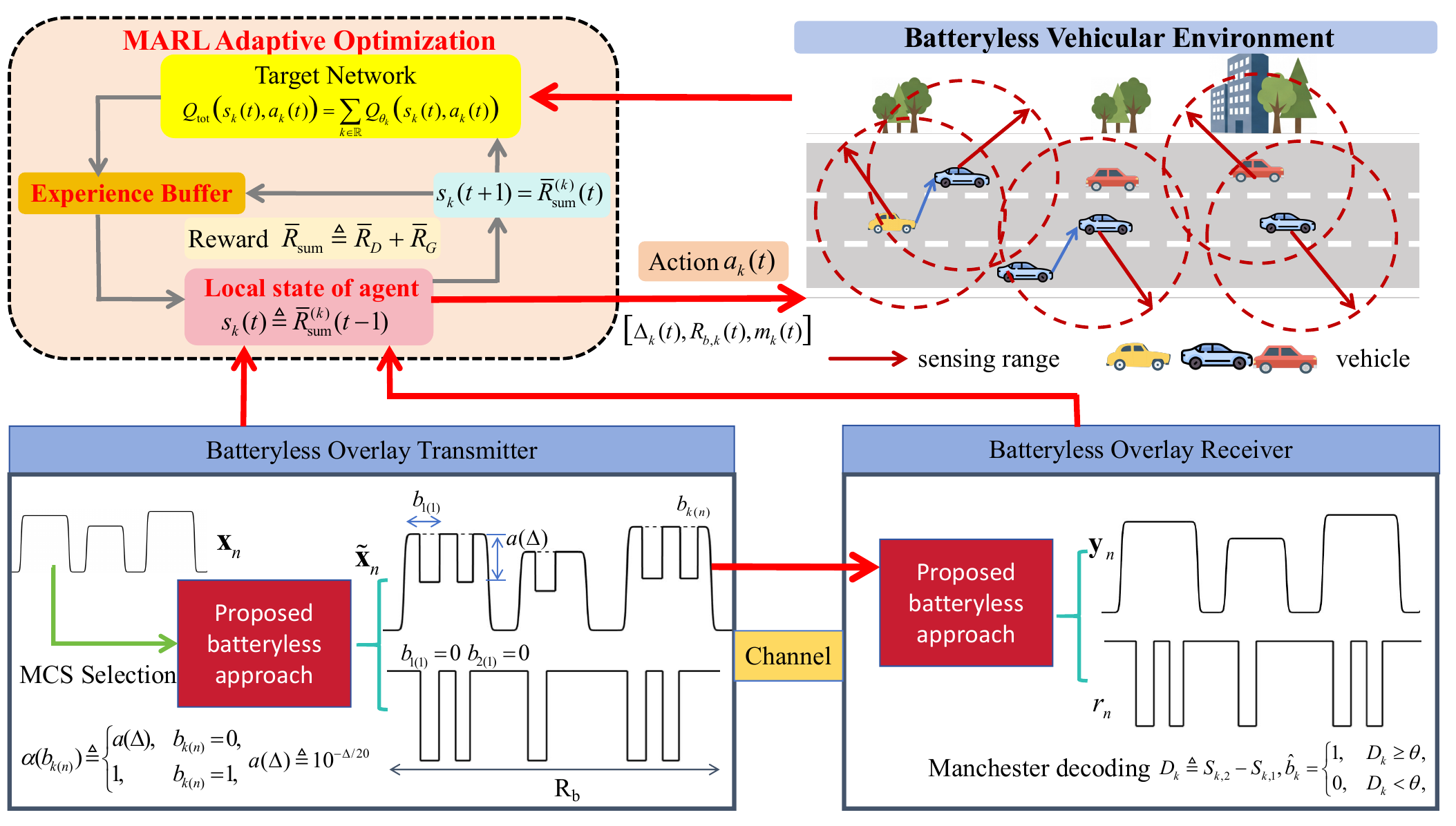}
    \caption{Batteryless V2X System Architecture. }
    \label{fig:system}
    \vspace{-0.7cm}
\end{figure*}

\section{System Architecture and System Model}
\subsection{System Architecture}

As Fig.~\ref{fig:system} shows, We consider a vehicular transmission system in which a legacy transmitter simultaneously serves a conventional vehicular receiver and a batteryless receiver. On the transmitter side, the legacy vehicular node first generates a standard DSRC packetized waveform according to the selected MCS. This host waveform then passes through a batteryless overlay module, which embeds a low-rate passive payload by applying controlled amplitude attenuation. The attenuation depth is determined by $\Delta$, while the embedding rate is determined by $R_b$.

The composite signal is transmitted over the vehicular channel and is received by two types of receivers. The legacy receiver decodes the host packet using the conventional PHY processing chain, treating the overlay-induced attenuation as tolerable distortion. In contrast, the batteryless receiver employs a low-power envelope detector to extract the amplitude variations and then recovers the embedded bits through Manchester decoding. In this way, the same packet transmission simultaneously supports legacy communication and passive downlink delivery. To adapt to different channel conditions, the architecture further includes an MARL-based PHY controller. Based on historical LUT-based PHY evaluation, the controller selects the control tuple. We propose an architecture that jointly integrates the batteryless overlay transmitter, the low-complexity passive receiver, and the adaptive PHY controller into a unified vehicular communication system.

\subsection{System Model}

Let $x_n(t)$ denote the host waveform of a packet transmission. The batteryless overlay bit stream is Manchester encoding, and the embedded-bit rate is denoted by $R_b$, with bit duration $T_b=1/R_b$. We associate each host packet $n$ with one overlay bit $b_{k(n)}\in\{0,1\}$, where $k(n)$ denotes the index of the embedding interval associated with packet $n$. Then, the transmitted waveform during packet $n$ is written as
\begin{equation}
s_n(t)=a\!\left(b_{k(n)}\right)x_n(t),
\label{eq:overlay_waveform}
\end{equation}
where the amplitude factor is defined as
\begin{equation}
a\!\left(b_{k(n)}\right)=
\begin{cases}
a(\Delta)=10^{-\Delta/20}, & b_{k(n)}=0,\\
1, & b_{k(n)}=1.
\end{cases}
\label{eq:amp_factor}
\end{equation}
where $\Delta$ denotes the attenuation depth in dB. A larger $\Delta$ produces a stronger passive signature, but also causes a larger perturbation to the host waveform.

For a packetized MIMO-OFDM transmission, we adopt a packet-level effective SINR abstraction. Let the host baseband vector on packet $n$ be
\begin{equation}
\mathbf{x}_n=\sqrt{P_n}\mathbf{V}_n\mathbf{s}_n,
\label{eq:host_packet}
\end{equation}
where $P_n$ is the packet power, $\mathbf{V}_n$ is the precoder, and $\mathbf{s}_n$ is the spatial-stream symbol vector with $\mathbb{E}[\mathbf{s}_n\mathbf{s}_n^H]=\mathbf{I}$. To preserve the spatial structure of the host packet, the proposed overlay applies a common scalar attenuation across all transmit antennas. Therefore, the transmitted vector becomes
\begin{equation}
\tilde{\mathbf{x}}_n=a\!\left(b_{k(n)}\right)\mathbf{x}_n,
\label{eq:overlay_packet}
\end{equation}
At the legacy receiver, the received signal is
\begin{equation}
\mathbf{y}_n=\mathbf{H}_n\tilde{\mathbf{x}}_n+\mathbf{w}_n,
\label{eq:legacy_rx}
\end{equation}
where $\mathbf{H}_n$ is the MIMO channel matrix and $\mathbf{w}_n\sim\mathcal{CN}(0,N_0\mathbf{I})$ is the receiver noise. The overlay acts as a controlled distortion term. Using a post-equalization abstraction, the effective SINR of packet $n$ is modeled as
\begin{equation}
\gamma_n^{\mathrm{W}}
=
\frac{
P_n\,\mathbb{E}\!\left[\left\|\mathbf{U}_n^H\mathbf{H}_n\mathbf{V}_n\right\|_F^2\right]
}{
N_0+\sigma_{\mathrm{ovl}}^2(\Delta,R_b;\mathbf{H}_n)
},
\label{eq:legacy_sinr}
\end{equation}
where $\mathbf{U}_n$ is the linear equalizer and $\sigma_{\mathrm{ovl}}^2(\Delta,R_b;\mathbf{H}_n)$ captures the overlay-induced impairment. The corresponding legacy packet error rate is represented by a PHY abstraction function
\begin{equation}
P_{e,\mathrm{W}}(n)=f_{\mathrm{W}}\!\left(\gamma_n^{\mathrm{W}},m\right),
\label{eq:legacy_per}
\end{equation}
where $m$ is the selected legacy MCS.

At the batteryless receiver, coherent demodulation is avoided. Let
\begin{equation}
r_n = g_n \tilde{\mathbf{x}}_n + v_n
\label{eq:passive_rx}
\end{equation}
denote the received scalar signal, where $g_n$ is the effective channel gain and $v_n$ is the receiver noise. After envelope detection, the observation is modeled as
\begin{equation}
z_n=|r_n|+\eta_n,
\label{eq:envelope_obs}
\end{equation}
where $\eta_n$ denotes residual detector noise. For Manchester-compatible decoding, the sample set associated with the $k$-th embedding interval is divided into two halves, denoted by $\mathcal{I}_{k,1}$ and $\mathcal{I}_{k,2}$, respectively. The corresponding average envelope observations are defined as
\begin{equation}
S_{k,1}=\frac{1}{|\mathcal{I}_{k,1}|}\sum_{n\in\mathcal{I}_{k,1}} z_n,
\end{equation}
\begin{equation}
S_{k,2}=\frac{1}{|\mathcal{I}_{k,2}|}\sum_{n\in\mathcal{I}_{k,2}} z_n.
\label{eq:avg_stat}
\end{equation}
where $S_{k,1}$ and $S_{k,2}$ denote the average envelope levels over the first and second halves of the $k$-th Manchester-coded bit interval, respectively. Therefore, the amplitude transition within one embedded bit period is captured by
\begin{equation}
D_k=S_{k,2}-S_{k,1}.
\label{eq:decision_stat}
\end{equation}
The detected bit is then written as
\begin{equation}
\hat{b}_k=
\begin{cases}
1, & D_k\ge 0,\\
0, & D_k<0.
\end{cases}
\label{eq:bit_decision}
\end{equation}
In this notation, $\hat{b}_k$ denotes the detected overlay bit in the $k$-th embedding interval, while $b_{k(n)}$ denotes the transmitted bit. Accordingly, the passive-link packet error probability is denoted by $P_{e,\mathrm{G}}$.

In addition to decoding reliability, the passive transmission must also satisfy a packet-duration feasibility requirement, since the overlay is carried only within the host packet. Let $T_{\mathrm{data}}(n)$ denote the available embedding time in packet $n$, and let $T_{\mathrm{pre}}$ and $K_{\mathrm{coded}}$ denote the preamble duration and the number of coded overlay bits per passive frame, respectively. Then, the embedding feasibility condition is given by
\begin{equation}
T_{\mathrm{pre}}+\frac{K_{\mathrm{coded}}}{R_b}\le T_{\mathrm{data}}(n),\forall n.
\label{eq:fit_condition}
\end{equation}

This condition shows that the embedded-bit rate is constrained not only by passive-link decoding performance, but also by the available host-packet duration. Together with the host-link and passive-link reliability metrics defined above, Eq. \eqref{eq:fit_condition} completes the PHY-layer model used in the subsequent sum-throughput optimization.

\section{Problem Formulation and Optimization}
\subsection{Problem Formulation}
In this section, we formulate the PHY-layer cooperative sum-throughput maximization problem for the proposed batteryless overlay approach. Our optimization concentrates on how to select the embedding and legacy PHY parameters such that the passive payload can be delivered while the quality of the host link remains acceptable. 
Let $T_{\mathrm{obs}}$ denote the observation horizon. For the $i$-th passive frame embedded into host packet $n$, let $K_{\mathrm{pay}}^{(n,i)}$ denote the net passive payload. Let $P_{\mathrm{det}}(\tau)=1-P_{\mathrm{MD}}(\tau)$ denote the
packet-detection success probability of the passive receiver,
where $\tau$ is the detection threshold and $P_{\mathrm{MD}}(\tau)$
is the corresponding missed-detection probability. The term
$P_{\mathrm{det}}(\tau)$ captures whether the passive receiver can
successfully detect the presence of an embedded passive frame
before decoding its payload. Let
$P^{G}_{e,\mathrm{dec}}(\Delta,T_b,\theta;K^{(n,i)}_{\mathrm{pay}})$
denote the passive-link decoding error probability conditioned
on successful detection and synchronization. Then, the net
passive-link throughput can be expressed as
\begin{equation}
\bar{R}_{\mathrm{G}}
=
\frac{1}{T_{\mathrm{obs}}}
\mathbb{E}
\!\left[
\sum_{n}
\sum_{i}
K_{\mathrm{pay}}^{(n,i)}
P_{\mathrm{det}}(\tau)
\Bigl(
1-
P_{e,\mathrm{dec}}^{\mathrm{G}}
\Bigr)
\right],
\label{eq:RB_general_conf}
\end{equation}

The legacy-link throughput is similarly determined by the
post-equalization SINR, the selected modulation and coding
scheme (MCS), and the embedding-induced distortion. Therefore,
the PHY-layer cooperative objective can be generally written as
\begin{equation}
\bar{R}_{\mathrm{sum}} = \bar{R}_{D} + \bar{R}_{G},
\vspace{-0.2cm}
\end{equation}
where $\bar{R}_{D}$ is the average legacy throughput under the
proposed batteryless overlay approach and $\bar{R}_{G}$ is the
average throughput of the passive link. This objective directly
reflects the essential tradeoff behind the proposed batteryless
overlay: the overlay is beneficial only when the additional passive
payload does not excessively degrade the host-link throughput.

Under PHY-layer adaptation, the controllable variables are
the attenuation depth $\Delta$, the embedded-bit rate $R_b$,
and the legacy MCS index $m$. For a different SINR environment
indexed by $k$, the PHY evaluation module, implemented by a
PHY simulator or an LUT, returns the legacy-link packet error
rate and passive-link packet error rate as $\left(P_{e,D}^{(k)}, P_{e,G}^{(k)}\right)
=
\mathrm{PHYEval}(k,\Delta,R_b,m)$. Let $R_D(m)$ denote the nominal host-link rate associated with
the selected MCS. Then, the general objective in (16) is
instantiated under environment $k$ as
\begin{equation}
\bar{R}_{\mathrm{sum}}^{(k)}
=
R_D(m)\left(1-P_{e,D}^{(k)}\right)
+
R_b\left(1-P_{e,G}^{(k)}\right).
\end{equation}
where the first term represents the effective host-link throughput
after accounting for legacy-link packet errors, while the second
term represents the effective passive-link throughput after accounting
for passive-link packet errors. To guarantee reliable communication, we impose packet-error-rate constraints 
on both links. Let $\epsilon_{\mathrm{D}}$ denote the maximum tolerable PER 
for the legacy host link, and let $\epsilon_{\mathrm{G}}$ denote the maximum 
tolerable PER for the batteryless passive link. The host-link reliability and 
passive-link reliability requirements are then expressed as
\vspace{-0.1cm}
\begin{equation}
P_{e,\mathrm{D}}^{(k)} \le \epsilon_{\mathrm{D}},\quad
P_{e,\mathrm{G}}^{(k)} \le \epsilon_{\mathrm{G}},\quad \forall k.
\label{eq:per_constraints}
\end{equation}

The problem is to achieve maximum throughput, but the PHY parameters vary 
due to environmental changes. Therefore, we obtain the following objective 
problem and constraint conditions.
\begin{equation*}
\begin{aligned}
&\text{\textbf{Problem 1: Cooperative Sum-Throughput Maximization}}\\
&\underset{\Delta,R_b,m}{\mathrm{argmin}}\quad
\bar{R}_{\mathrm{sum}}^{(k)}(\Delta,R_b,m) \\
&\mathrm{s.t.}\quad \mathrm{C1}:~
P_{e,\mathrm{D}}^{(k)}(\Delta,R_b,m)\le \epsilon_{\mathrm{D}},\ \forall k,\\
&\phantom{\mathrm{s.t.}\quad}\mathrm{C2}:~
P_{e,\mathrm{G}}^{(k)}(\Delta,R_b,m)\le \epsilon_{\mathrm{G}},\ \forall k,\\
&\phantom{\mathrm{s.t.}\quad}\mathrm{C3}:~
T_{\mathrm{pre}}+\frac{K_{\mathrm{coded}}}{R_b}\le T_{\mathrm{data}}^{(k)}(m),\ \forall k,\\
&\phantom{\mathrm{s.t.}\quad}\mathrm{C4}:~
0\le \Delta \le \Delta_{\max},
R_b\in \mathcal{A}_{R_b},
m\in \mathcal{A}_{\mathrm{MCS}}.
\end{aligned}
\label{eq:problem1_conf}
\end{equation*}
\textbf{Problem 1} highlights three coupled PHY tradeoffs: A larger attenuation depth improves passive detectability but increases host-link distortion; a larger $R_b$ enlarges the nominal passive throughput but makes passive decoding more difficult; and a larger MCS improves the nominal host rate but may shorten the available embedding duration within the packet. As a result, the host-link and passive-link objectives are intrinsically coupled through the same control tuple $(\Delta,R_b,m)$ and cannot be optimized separately.

\subsection{MARL for Adaptive Optimization}

Problem 1 is difficult to solve analytically because the mapping from the PHY control tuple $(\Delta,R_b,m)$ to the host-link reliability, passive-link reliability, and the resulting sum-throughput is not available in closed form. Instead, these quantities are obtained numerically from LUTs or PHY simulations. Therefore, rather than pursuing repeated analytical optimization, we adopt a lightweight multi-agent reinforcement learning (MARL) solver to search the discrete PHY action space efficiently. We consider $|\mathcal{K}|$ agents, each associated with one SINR environment $k\in\mathcal{K}$. The $k$-th agent learns a PHY parameter-selection policy tailored to $\mathcal{E}^{(k)}$.
In implementation, $\mathcal{E}^{(k)}$ can be instantiated by a SINR-specific dataset. At decision step $t$, agent $k$ selects the PHY control tuple
\begin{equation}
a_k(t) = [\Delta_k(t),\,R_{b,k}(t),\,m_k(t)],
\label{eq:action_tuple_conf}
\end{equation}
from the discrete feasible set $\mathcal{A}= \mathcal{A}_{\Delta}\times \mathcal{A}_{R_b}\times \mathcal{A}_{\mathrm{MCS}}.
\label{eq:action_space_conf2}$ For a chosen action $a_k(t)$, the LUT-based PHY evaluation module returns the corresponding PER pair $\bigl(P_{e,\mathrm{D}}^{(k)}(t),\,P_{e,\mathrm{G}}^{(k)}(t)\bigr)
=
\mathrm{PHYEval}\bigl(k,a_k(t)\bigr)$,
which yields the per-environment PHY-layer throughput
% \begin{equation}
% \hat R_{\mathrm{sum}}^{(k)}(t)
% =
% R_{\mathrm{W}}\!\bigl(m_k(t)\bigr)\bigl(1-P_{e,\mathrm{W}}^{(k)}(t)\bigr)
% +
% R_{b,k}(t)\bigl(1-P_{e,\mathrm{B}}^{(k)}(t)\bigr).
% \label{eq:sum_goodput}
% \end{equation}
\begin{equation}
\small
\begin{aligned}
\bar R_{\mathrm{sum}}^{(k)}(t)
={}&
R_{\mathrm{D}}\!\bigl(m_k(t)\bigr)\bigl(1-P_{e,\mathrm{D}}^{(k)}(t)\bigr)
+
R_{b,k}(t)\bigl(1-P_{e,\mathrm{G}}^{(k)}(t)\bigr),
\end{aligned}
\label{eq:sum_goodput}
\end{equation}

Following the throughput-oriented design, we choose the local state of agent $k$ as the previous-step measured sum-throughput,
\begin{equation}
s_k(t)= \bar R_{\mathrm{sum}}^{(k)}(t-1).
\label{eq:local_state_conf}
\end{equation}
The shared objective across the environments is defined through the average sum-throughput
\begin{equation}
\bar{R}_{\mathrm{sum}}(t) 
= \frac{1}{|\mathcal{K}|}\sum_{k\in\mathcal{K}} \bar{R}_{\mathrm{sum}}^{(k)}(t),
\label{eq:avg_sum_goodput_conf}
\end{equation}
To favor high-throughput yet feasible operating points, we define the reward as
\begin{equation}
\mathcal{R}(t)=\exp\!\bigl(\eta \bar R_{\mathrm{sum}}(t)\bigr)-\lambda M^2(t),
\label{eq:reward}
\end{equation}
where $\eta>0$ is a scaling factor, $\lambda>0$ is the penalty weight, and $M(t)$ denotes the number of fixed-SINR environments violating the PER constraints in \textbf{Problem~1}. In this way, the first term encourages large cooperative throughput, whereas the second term suppresses infeasible PHY configurations.

\begin{algorithm}[t]
\footnotesize
\caption{MARL for PHY Parameter Adaptation}
\label{alg:dpfvd}
\DontPrintSemicolon
\KwIn{Fixed-SINR set $\mathcal{K}=\mathbb{R}$ dB; action space $\mathcal{A}=\mathcal{A}_{\Delta}\times\mathcal{A}_{R_b}\times\mathcal{A}_{\mathrm{MCS}}$; reward parameters $(\eta,\lambda)$; PER thresholds $(\epsilon_{\mathrm{D}},\epsilon_{\mathrm{G}})$.}
\KwOut{Learned local policies $\{\pi^{(k)}\}_{k\in\mathcal{K}}$.}

Initialize local Q-networks $\{Q_{\theta_k}\}_{k\in\mathcal{K}}$, target networks $\{\bar Q_{\theta_k}\}_{k\in\mathcal{K}}$, and replay buffer $\mathcal{D}$\;

\For{each episode}{
    Initialize local states $\{s_k(1)\}_{k\in\mathcal{K}}$\;
    \For{$t=1,2,\ldots,T$}{
        \ForEach{$k\in\mathcal{K}$}{
            Select $a_k(t)=[\Delta_k(t),R_{b,k}(t),m_k(t)]$ via $\epsilon$-greedy\;
            Query PHY:
            $(P_{e,\mathrm{D}}^{(k)},P_{e,\mathrm{G}}^{(k)})=\mathrm{PHYEval}(k,a_k(t))$\;
            Compute $\bar R_{\mathrm{sum}}^{(k)}(t)$ via Eq. \eqref{eq:sum_goodput}\;
            Update $s_k(t+1)\leftarrow \bar R_{\mathrm{sum}}^{(k)}(t)$\;
        }
        Compute $\bar R_{\mathrm{sum}}(t)$ and reward $\mathcal{R}(t)$ via Eq. \eqref{eq:reward}\;
        Store $(s(t),a(t),\mathcal{R}(t),s(t+1))$ in $\mathcal{D}$\;
        Update $\{Q_{\theta_k}\}$ using \eqref{eq:td_loss}\;
        Copy $\{Q_{\theta_k}\}\rightarrow\{\bar Q_{\theta_k}\}$ periodically\;
    }
}
\end{algorithm}

To enable cooperative learning across the three environments, we adopt a value-decomposition structure. Each agent maintains a local action-value function $Q_{\theta_k}(s_k,a_k)$, and the global $Q$-function is formed as
\begin{equation}
Q_{\mathrm{tot}}(s(t),a(t))
=
\sum_{k\in\mathbb{R}}Q_{\theta_k}\bigl(s_k(t),a_k(t)\bigr),
\label{eq:qtot}
\end{equation}

Given the transition tuple $(s(t),a(t),\mathcal{R}(t),s(t+1))$, the local networks are trained by minimizing the temporal-difference loss on the decomposed global $Q$-function:
% \begin{equation}
% J
% =
% \mathbb{E}\!\left[
% \left(
% Q_{\mathrm{tot}}(s(t),a(t))
% -
% \Bigl(
% \mathcal{R}(t)+\gamma \max_{a'}Q_{\mathrm{tot}}(s(t+1),a')
% \Bigr)
% \right)^2
% \right].
% \label{eq:td_loss}
% \end{equation}

\vspace{-0.5cm}
\begin{equation}
\small
\begin{aligned}
J
=&\,
\mathbb{E}\!\Biggl[
\Bigl(
Q_{\mathrm{tot}}(s(t),a(t))
-
\bigl(
\mathcal{R}(t)+\gamma \max_{a'}Q_{\mathrm{tot}}(s(t+1),a')
\bigr)
\Bigr)^2
\Biggr].
\end{aligned}
\label{eq:td_loss}
\end{equation}

 Since the expensive PHY evaluation is already moved offline into LUT construction, online adaptation only needs to search a finite action set and update the policy according to the observed LUT outputs. Therefore, MARL here serves as an efficient solver for Problem 1, rather than a separate system layer. The detailed procedure regarding PHY Parameters adaptation using MARL is summarized in \textbf{Algorithm 1}.

\begin{figure*}[!t]
  \centering

  % ===================== 第一行：DSRC =====================
  \setcounter{subfigure}{0} % 第一行从 (a) 开始

  \subfloat[DSRC PER under Different $\Delta$.]{%
    \includegraphics[width=0.32\textwidth]{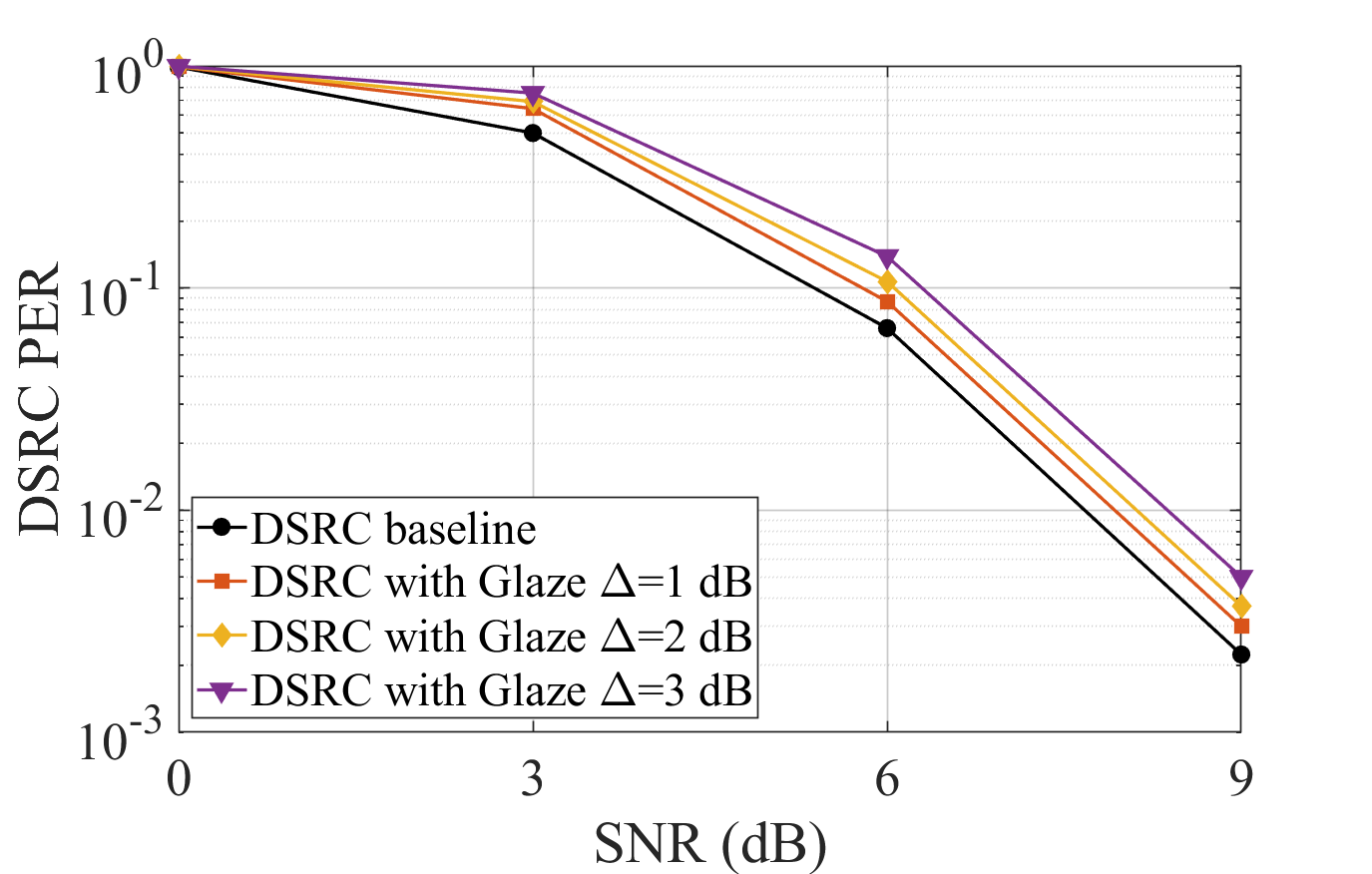}%
    \label{fig:wifi_delta}}
  \hfill
  \subfloat[DSRC PER under Different MCS.]{%
    \includegraphics[width=0.32\textwidth]{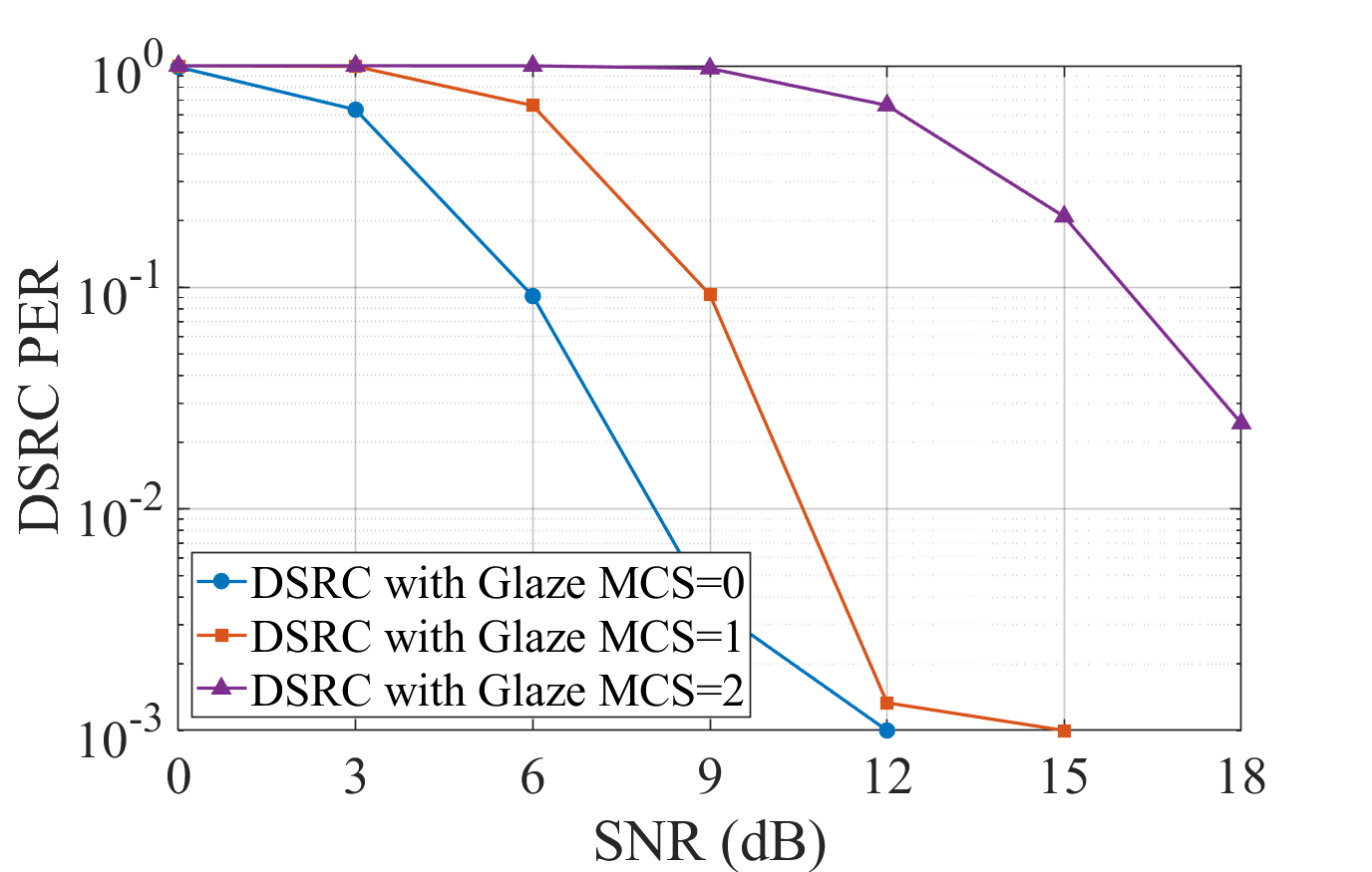}%
    \label{fig:wifi_mcs}}
  \hfill
  \subfloat[DSRC PER under Different $R_b$.]{%
    \includegraphics[width=0.32\textwidth]{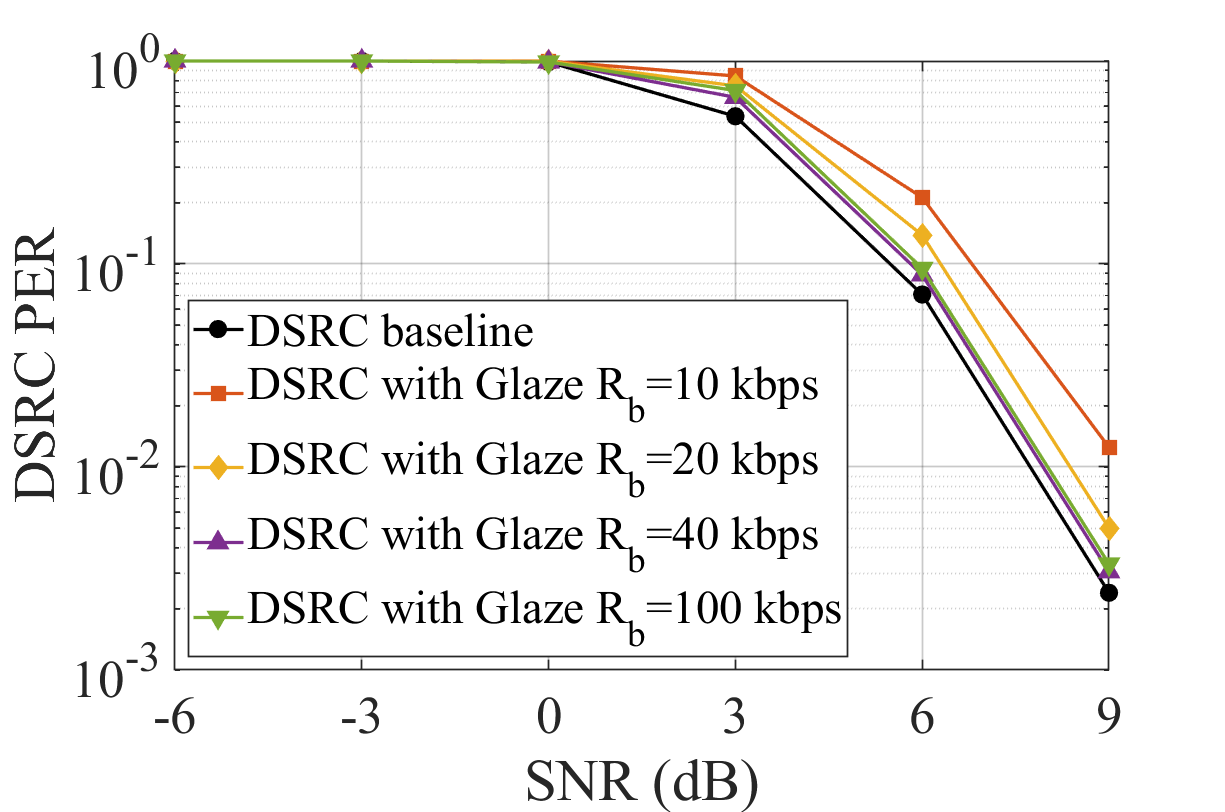}%
    \label{fig:wifi_rb}}

\vspace{-0.2cm}

  \captionof*{figure}{Fig.~2: DSRC with Glaze System: DSRC PER Performance in terms of PHY Parameters.}
  \label{fig:dsrc_per_summary}

  % ===================== 第二行：Glaze =====================
  \setcounter{subfigure}{0} % 第二行重新从 (a) 开始
\vspace{-0.3cm}

  \subfloat[Glaze PER under Different $\Delta$.]{%
    \includegraphics[width=0.32\textwidth]{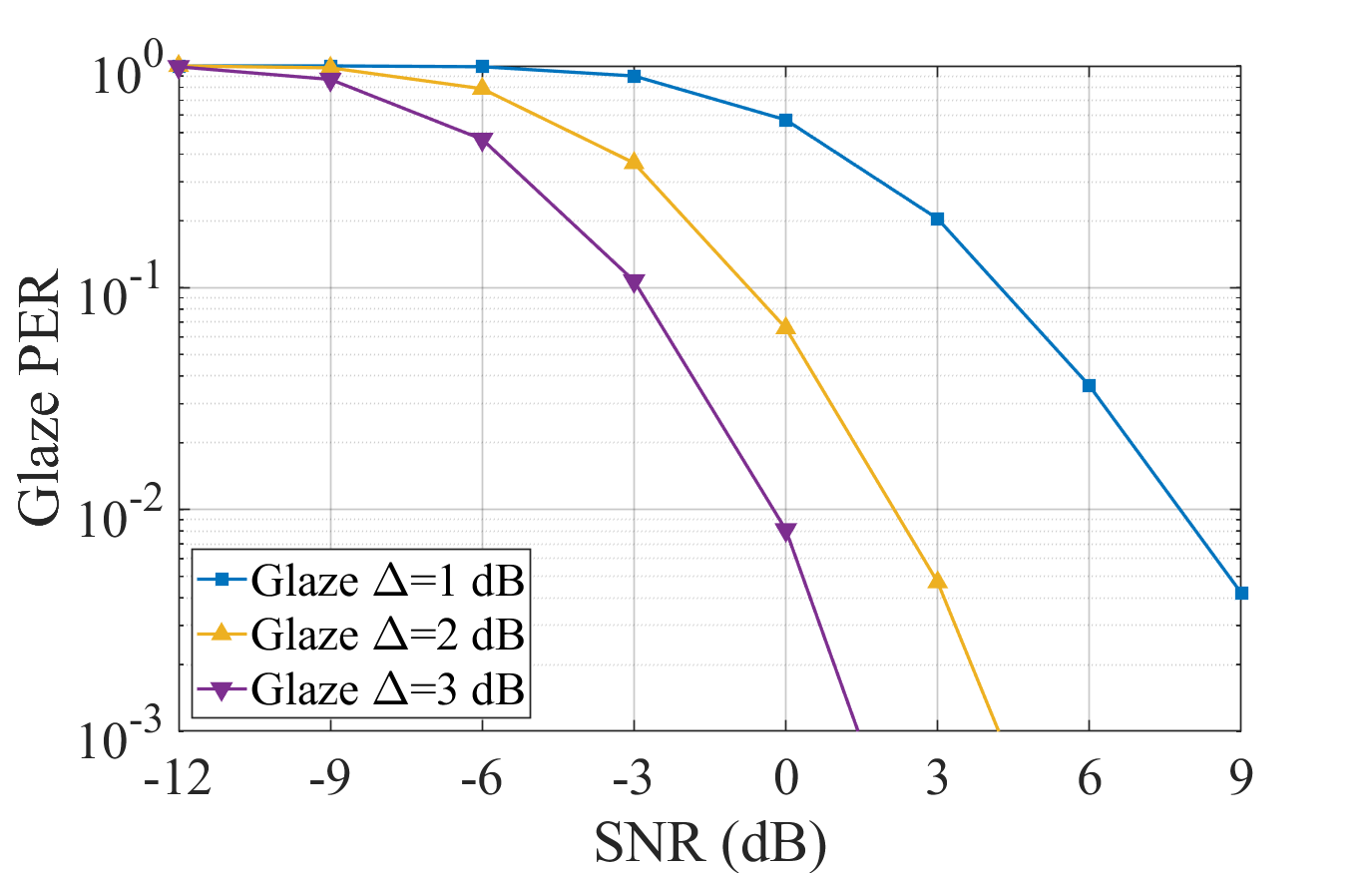}%
    \label{fig:glaze_delta}}
  \hfill
  \subfloat[Glaze PER under Different MCS.]{%
    \includegraphics[width=0.32\textwidth]{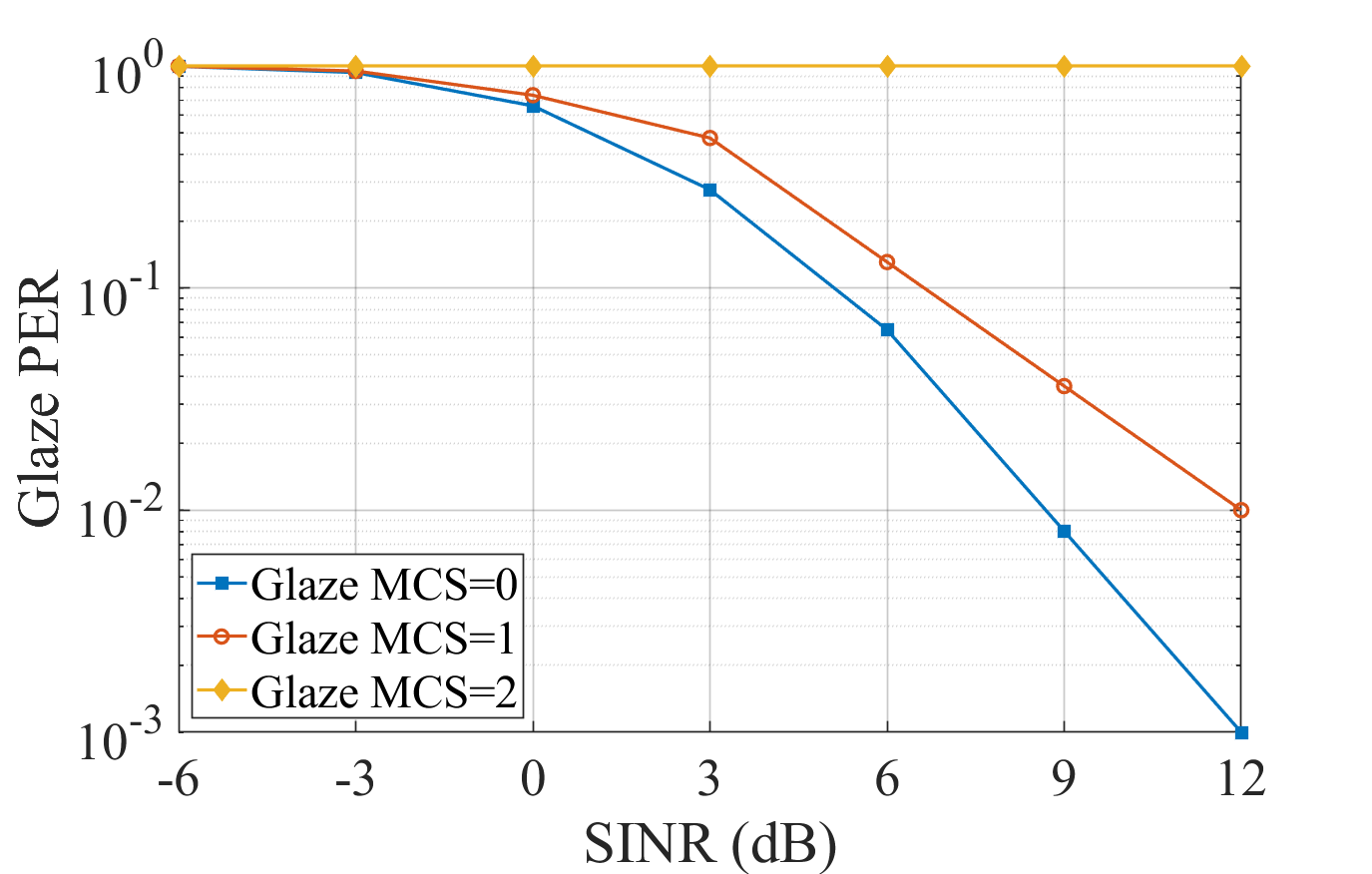}%
    \label{fig:glaze_mcs}}
  \hfill
  \subfloat[Glaze PER under Different $R_b$.]{%
    \includegraphics[width=0.32\textwidth]{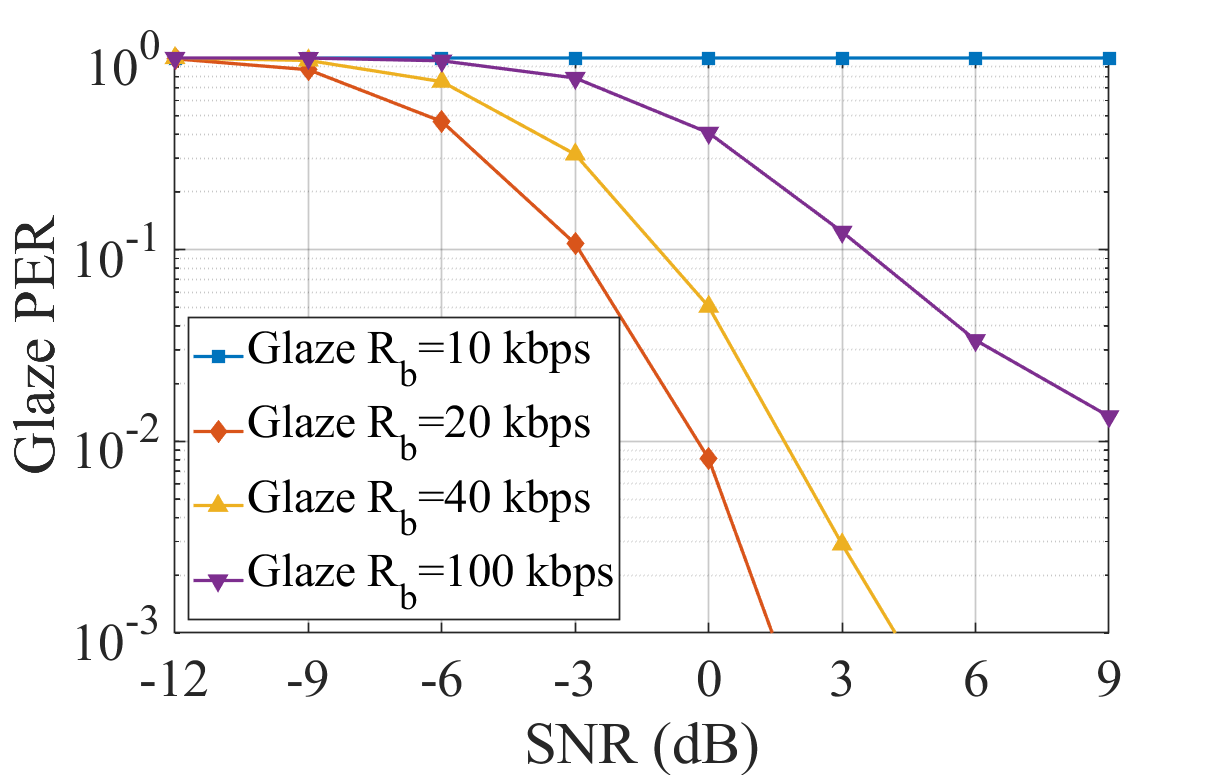}%
    \label{fig:glaze_rb}}
\vspace{-0.2cm}
  \captionof*{figure}{Fig.~3: DSRC with Glaze System: Glaze PER Performance in terms of PHY Parameters.}
  \label{fig:glaze_per_summary}
\vspace{-0.5cm}
\end{figure*}

\section{Simulation}
In this section, we evaluate the PHY-layer impact of the three control variables, namely the attenuation depth $\Delta$, the embedded-bit rate $R_b$, and the legacy MCS. The results are obtained under the LUT-based PHY evaluation framework described in the previous sections. MATLAB is employed to perform PHY-layer waveform simulation and PER analysis, whereas Python is used to implement the MARL-based joint optimization for adaptive parameter selection. Since this conference version focuses on PHY-only adaptation, we emphasize the coupled effects of these variables on the host-link PER and the passive-link PER. The main simulation parameters are summarized in Table I.
\begin{table}[t]
\centering
\caption{SIMULATION PARAMETERS}
\label{tab:sim_params}
\resizebox{0.7\columnwidth}{!}{
\begin{tabular}{|l|l|}
\hline
\textbf{Parameters} & \textbf{Value} \\
\hline
Legacy V2X/Wi-Fi MCS & \{0, 1, 2\} \\
\hline
Legacy packet length & 1000 bits \\
\hline
Carrier frequency & 5.9 GHz \\
\hline
AWGN power spectral density $N_0$ & -114 dBm \\
\hline
Legacy transmit power & 23 dBm \\
\hline
Attenuation depth $\Delta$ & \{1, 2, 3, 5\} dB \\
\hline
Embedded bit rate $R_b$ & \{10, 20, 40, 100\} kbps \\
\hline
Passive receiver sampling rate & 1 MHz \\
\hline
Envelope detector sensitivity & -60 dBm \\
\hline
SNR range & -15 to 23 dB \\
\hline
Simulation runs per SNR & 10000 \\
\hline
\end{tabular}}
\end{table}

Fig.~2 shows the DSRC PER performance under varying PHY parameters, while Fig.~3 presents the corresponding Glaze PER performance. As shown in Fig.~2(a), the host-link PER remains close to the DSRC baseline when $\Delta$ is small, while a larger $\Delta$ gradually degrades the host-link reliability, especially at high SNR. In contrast, Fig.~3(a) shows that the passive-link PER is very high when $\Delta$ is too small, because the embedded amplitude variation is difficult to distinguish. As $\Delta$ increases, the passive-link PER decreases significantly. Therefore, $\Delta$ directly determines the fundamental PHY tradeoff: a small $\Delta$ protects the host packet, whereas a large $\Delta$ improves passive-link decodability. As shown in Fig.~2(b), lower MCS values provide better host-link robustness, while higher MCS values become increasingly sensitive to SNR and overlay-induced distortion. Fig.~3(b) further shows that lower MCS values also benefit passive reception. The main reason is that a lower MCS yields a longer host-packet duration, which provides more time for passive embedding and detection. By contrast, when the MCS becomes too high, the packet duration shortens and the passive transmission becomes unreliable or even infeasible. Hence, the legacy MCS should be optimized jointly with the overlay parameters rather than selected independently. As shown in Fig.~2(c), the host-link PER changes with $R_b$, indicating that the embedded-bit rate also affects the disturbance imposed on the host waveform. Fig.~3(c) shows that passive reception is also highly sensitive to $R_b$. If $R_b$ is too large, the observation time per embedded bit becomes too short, which increases the passive-link PER. If $R_b$ is too small, the required embedding duration becomes too long relative to the available packet time, which is also unfavorable for passive delivery. Therefore, $R_b$ should be selected jointly with $\Delta$ and the legacy MCS.

% Overall, Figs.~2--4 show that the proposed batteryless overlay cannot be optimized by tuning a single PHY parameter in isolation. The best operating point must jointly balance host-link reliability, passive-link reliability, and packet-duration feasibility. This observation directly motivates the PHY-only joint optimization problem and the adaptive MARL-based parameter selection proposed in this paper.
To further validate the effectiveness of the proposed MARL-based PHY adaptation, we show the training performance in Fig.~4. As shown in Fig.~4(a), the MARL training loss decreases gradually and then converges to a stable level, confirming the convergence and stability of the proposed optimization process. Fig.~4(b) shows that the average throughput increases from about $4.5$ Mbps to approximately $5.15$ Mbps during training, corresponding to an improvement of about $15\%$. This demonstrates that the proposed MARL controller can effectively optimize the PHY parameter tuple and achieve a higher throughput.

% \begin{figure}[h]
% \begin{minipage}[t]{0.48\linewidth}
% \centering
%  \subfigure[Throughput.]
%  {\includegraphics[width=1.7in]{figures/throughput.png}}
% \label{fig:4a}
% \end{minipage}
% \begin{minipage}[t]{0.24\linewidth}
% \centering
% \subfigure[Loss.]{
% \includegraphics[width=1.7in]{figures/loss.png}}
% \label{fig:4b}
% \end{minipage}%
% \caption{MARL.}
% \label{fig:4}
% \end{figure}
\setcounter{figure}{3}
\begin{figure}[t]
  \centering
  \subfloat[MARL Training Loss.]{%
    \includegraphics[width=0.8\columnwidth]{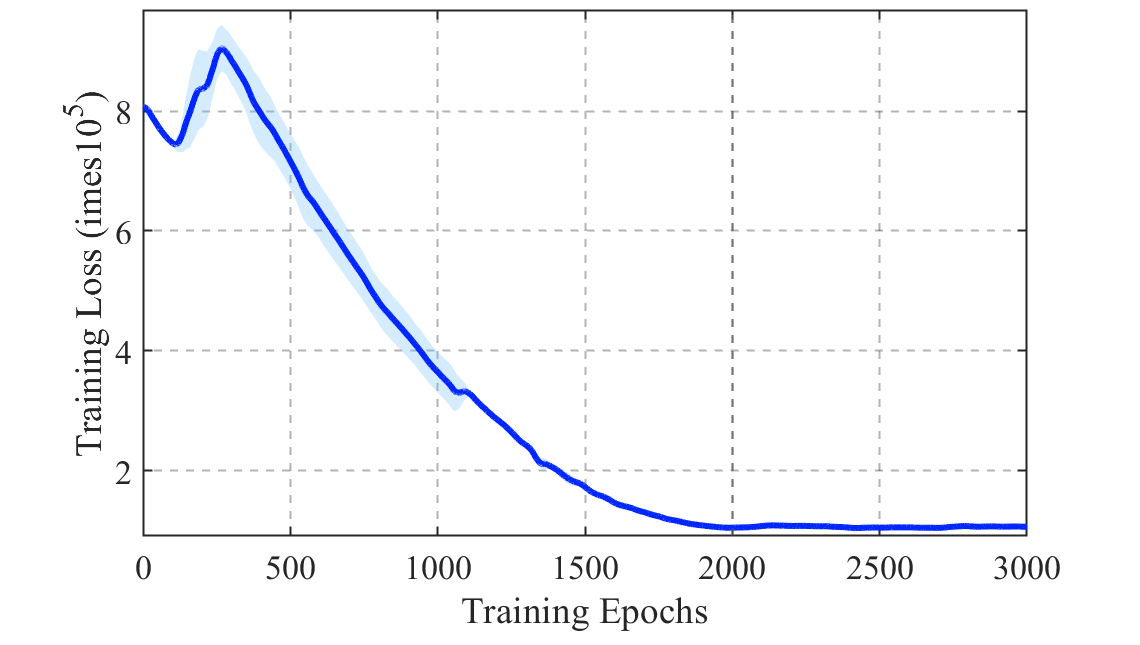}%
    \label{fig:Throughput}}
  \\[4pt]
  \subfloat[Throughput Comparison.]{%
    \includegraphics[width=0.8\columnwidth]{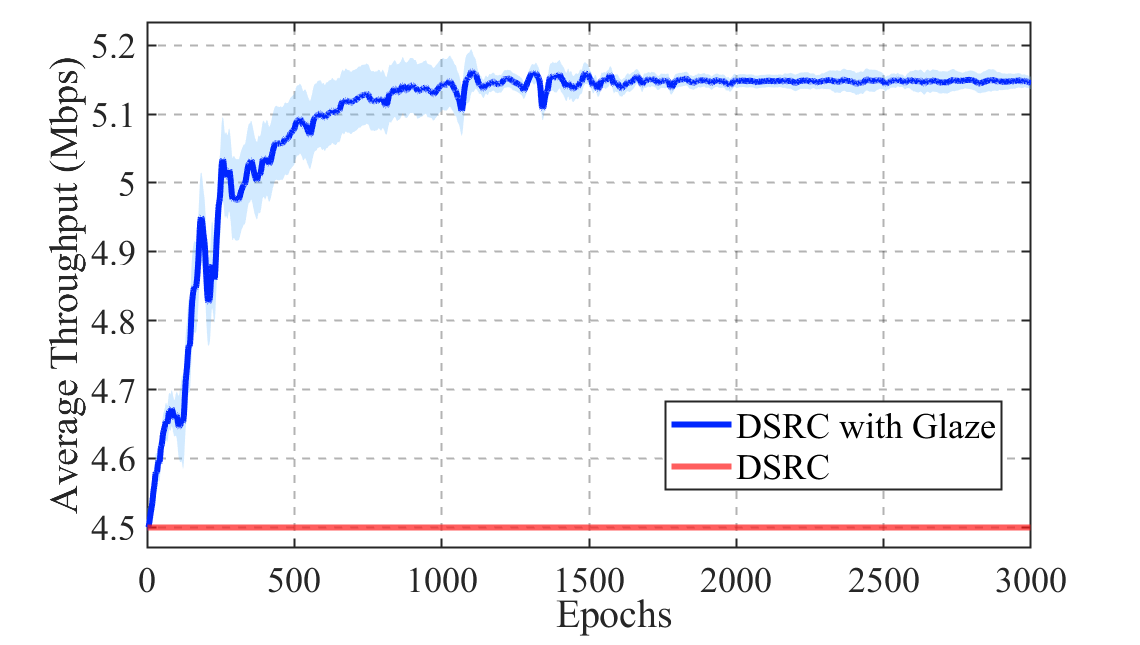}%
    \label{fig:Loss}}
  \caption{The MARL for Adaptive Optimization.}
  \label{fig:MARL}
\end{figure}
\section{Conclusion}
\label{sec:con}
In this paper, we proposed a PHY-layer batteryless V2X overlay framework, in which a legacy vehicular packet is reused to carry a passive downlink through controlled amplitude attenuation. Based on this model, we analyzed the coupled effects of the attenuation depth, embedded-bit rate, and legacy MCS on host-link decoding, passive-link decoding, and embedding feasibility, and formulated a constrained sum-throughput maximization problem. We further developed an MARL-based adaptive parameter-selection method over LUT-driven PHY evaluations.Simulation results showed that the attenuation depth, embedded-bit rate, and legacy MCS cannot be optimized independently, and that effective operating points must be obtained through joint PHY adaptation. In addition, the proposed MARL-based controller achieved stable convergence and improved the average throughput during training. Therefore, the proposed framework provides an effective solution for throughput-driven PHY adaptation in batteryless V2X overlay communications. Future work will extend the learned policy to more complete vehicular settings with richer channel and protocol dynamics.

\bibliographystyle{IEEEtran}
\bibliography{ref}
\end{document}